\documentclass[%
aps,
pre,
groupedaddress,
showkeys,
 amsmath,amssymb,
reprint,%
]{revtex4-2}
\usepackage{graphicx}
\usepackage{dcolumn}
\usepackage{bm}
\usepackage{xcolor}

\begin{document}


\title{A complexity measure
in natural time analysis identifying the accumulation of stresses before major earthquakes} 
\thanks{To the memory of the Academician Seiya Uyeda.}

\author{Panayiotis A. Varotsos}
\email{pvaro@otenet.gr}
\affiliation{Section of Condensed Matter Physics and Solid Earth Physics Institute, Physics Department, National and Kapodistrian University of Athens, Panepistimiopolis, Zografos 157 84, Athens, Greece}

\author{Nicholas V. Sarlis}
\affiliation{Section of Condensed Matter Physics and Solid Earth Physics Institute, Physics Department, National and Kapodistrian University of Athens, Panepistimiopolis, Zografos 157 84, Athens, Greece}

\author{Toshiyasu Nagao}
\affiliation{Natural Disaster Research Section (NaDiR), Global Center for Asian and Regional Research, University of Shizuoka, 3-6-1,Takajo, Aoi-Ku, Shizuoka-City, Shizuoka, 420-0839,Japan}



\date{\today}

\begin{abstract}
Here we suggest a new procedure through which one 
can identify when the accumulation of stresses before
 major earthquakes (EQs) (of magnitude $M$ 8.2 or larger) occurs. 
 By analyzing the seismicity in the frame of natural time, which is 
a new concept of time introduced in 2001,
 we study the evolution of the fluctuations of the entropy change
 of seismicity under time reversal for various scales of 
 different length  $i$  (number of events). We find that 
anomalous intersections between scales of
 different lengths $i$ are observed upon 
 approaching an extraordinary major EQ occurrence. The investigation 
 is presented for the seismicity in Japan 
since 1984 including the $M$9
Tohoku EQ  on 11 March 2011, which is the largest EQ ever recorded there.  
\end{abstract}
\maketitle 


\section{Introduction}
 It is widely known \cite{CAR94,HOL06,SPRINGER} that
earthquake (EQ) {occurrences} exhibit complex correlations in time, space and magnitude  (e.g., \cite{HUA08,HUA11,TEL09,LEN08,LEN11,RUN12}) 
and the observed
EQ scaling laws \cite{TUR97}  indicate the existence of phenomena closely associated with the
proximity of the system to a critical point.
In the 1980s,  the observation of 
 Seismic Electric Signals (SES), which are low frequency transient changes of the electric field of the Earth preceding
 EQs, was reported \cite{VAR84A,VAR84B,VAR86}. 
 Many SESs observed within a short time are termed SES activity \cite{VAR91} being accompanied by Earth's magnetic field variations \cite{VAR03} mainly  on the z-component \cite{SAR02,NEWBOOK}.
These observations have been motivated by a 
 physical model for SES generation, which 
 enables the explanation of the simultaneous 
 detection of  additional transient multidisciplinary
 phenomena before the EQ rupture\cite{angeo}.
This physical model  is termed ``pressure stimulated polarization currents'' (PSPC) model \cite{VARBOOK,VAR84A,VAR84B,VAR93} and could be summarized as follows:
 In the Earth, electric dipoles are always 
 present\cite{VARBOOK}  due to lattice imperfections 
 (point and linear defects) in the ionic constituents of rocks 
 and exhibit initially random orientations 
 at the future focal region of an EQ, where the stress, $\sigma$, starts to gradually increase. This is called stage A. 
 When this stress accumulation achieves a critical value, the electric dipoles exhibit a cooperative orientation 
 resulting in the emission of a  SES activity
 (cf.  cooperativity is 
 a hallmark of criticality \cite{STAN99}). This is called stage B. 
Uyeda et al. \cite{UYE09B} mentioned 
that the PSPC model is unique among other models
that have been proposed for the explanation of the SES 
generation. 

The criticality of SES activities has been ascertained by
 employing natural time analysis (NTA) 
\cite{NAT02,NAT03A,NAT03B},
 which has been introduced in 2001 \cite{NAT01} based on a new
 concept of time termed natural time. NTA 
 enables 
the uncovering of hidden properties in time
 series of complex systems and can identify 
 when the system approaches the critical point 
 {(for EQs the mainshock occurrence is considered the new phase)} 
 \cite{SPRINGER,SPRINGER23}.
 
\section{Natural time analysis. Background.}

For a time series comprising $N$ events, we define as natural time $\chi_k$  for the occurrence of the $k$-th event the quantity  $\chi_k=k/N$ \cite{NAT01,NAT02,NAT02A}. 
Hence, we ignore the time intervals between consecutive events, but preserve 
their order and energy $Q_k$. 
The evolution of the pair $(\chi_k,p_k)$ is studied, where  
$p_k=Q_k/\sum_{n=1}^N Q_n$
is the normalized energy for the  $k$-th event.
Using  $\Phi(\omega)=\sum_{k=1}^N p_k \exp (i \omega \chi_k)$ 
as the characteristic function of $p_k$ for all $\omega \in \mathcal {R}$, 
 the behavior of $\Phi(\omega)$ is studied at $\omega \rightarrow 0$,
 because all the moments 
of the distribution of $p_k$ can be estimated from 
the derivatives  $d^m \Phi(\omega) / d\omega^m$ (for $m$ positive integer) 
 at $\omega \rightarrow 0$. 
A quantity $\kappa_1$  was defined from the Taylor expansion $\Pi(\omega)= | \Phi(\omega) |^2 = 1- \kappa_1 \omega^2 + \kappa_2 \omega^4 + \ldots$ 
where 
\begin{equation}\kappa_1 = \langle
\chi^2 \rangle - \langle \chi \rangle^2 =\sum_{k=1}^N p_k (\chi_k)^2- \left(\sum_{k=1}^N p_k
\chi_k \right)^2. \label{k1}
\end{equation} 
A careful study  shows\cite{NAT05C}  that  $\kappa_1$  may be considered as an order parameter of seismicity and was also 
demonstrated\cite{PNAS15} that the spatiotemporal variations of $\kappa_1$ reveal
the epicenters of the EQs of 
magnitude $M\geq 7.6$.

The {\em dynamic} entropy $S$ in natural time is given by \cite{NAT03B}
\begin{equation}\label{eq3}
S=\langle \chi \ln \chi \rangle - \langle \chi \rangle \ln \langle \chi \rangle,
\end{equation} 
where $\langle f(\chi) \rangle=\sum_{k=1}^N p_k f(\chi_k)$ denotes the average value of $f(\chi)$ weighted by $p_k$, i.e.,
$\langle \chi \ln \chi \rangle = \sum_{k=1}^N p_k (k/N) \ln (k/N)$ and $\langle \chi \rangle = \sum_{k=1}^N p_k (k/N) $.
 Upon considering \cite{NAT05B,SPRINGER} the time-reversal $\hat{T}$, i.e., $\hat{T}p_k=p_{N-k+1}$, the entropy obtained by Eq. (\ref{eq3}), labelled by $S_-$, is given by  
\begin{eqnarray}\label{defS2}
S_-=\sum_{k=1}^N p_{N-k+1} \frac{k}{N} \ln \left( \frac{k}{N} \right)- \nonumber \\
\left(\sum_{k=1}^N p_{N-k+1} \frac{k}{N} \right) \ln \left( \sum_{k=1}^N p_{N-k+1} \frac{k}{N} \right), 
\end{eqnarray}
which  is different from $S$. 
Hence, there exists a change  $\Delta S \equiv S -S_-$ in natural
 time under time reversal, thus $S$ being  time-reversal 
 asymmetric \cite{NAT05B,NAT07,SPRINGER,SPRINGER23}. 
 The calculation of $\Delta S$ is carried out by means 
 of a window of length $i$ (=number of successive events), sliding each time by one event, through the whole time series, thus,  a new time series comprising successive $\Delta S_i$ values is formed.

The complexity measure $\Lambda_i$ is defined by \cite{SPRINGER,EPL15} 
\begin{equation}\label{eq5}
\Lambda_i = \frac{\sigma (\Delta S_i)}{\sigma (\Delta S_{100})}
\end{equation}
where $\sigma (\Delta S_i)$  is the standard deviation  of the time series of $\Delta S_i \equiv S_i -(S_-)_i$ and 
 the denominator stands for the standard 
 deviation $\sigma (\Delta S_{100})$ of the time series of $\Delta S_i$ of $i$=100 events. Thus, in short, 
 $\Lambda_i$ quantifies how the statistics of $\Delta S_i$ time 
 series varies upon changing the scale from 100 to another scale $i$, 
 and is of profound importance to study the dynamical evolution 
 of a complex system (see p. 159 of Ref.\cite{SPRINGER}).

\section{Results}  

We used the seismic catalog of the Japan Meteorological Agency (JMA) 
in a similar fashion 
as in Refs.\cite{PNAS13,PNAS15,VAR21} by
considering all EQs of magnitude $M\geq 3.5$ 
to assure data completeness from 1984 until 15 November 2023
 within the area $25^o - 46^o$N,
$125^o - 148^o$E. 
The EQ energy was obtained from the JMA magnitude $M$ by
 converting \cite{TAN04} to the 
 moment magnitude $M_w$ \cite{KAN78}.
The $\Lambda_i$ values were computed according to Eq.(\ref{eq5}).

\begin{figure}
\includegraphics[scale=0.35,angle=270]{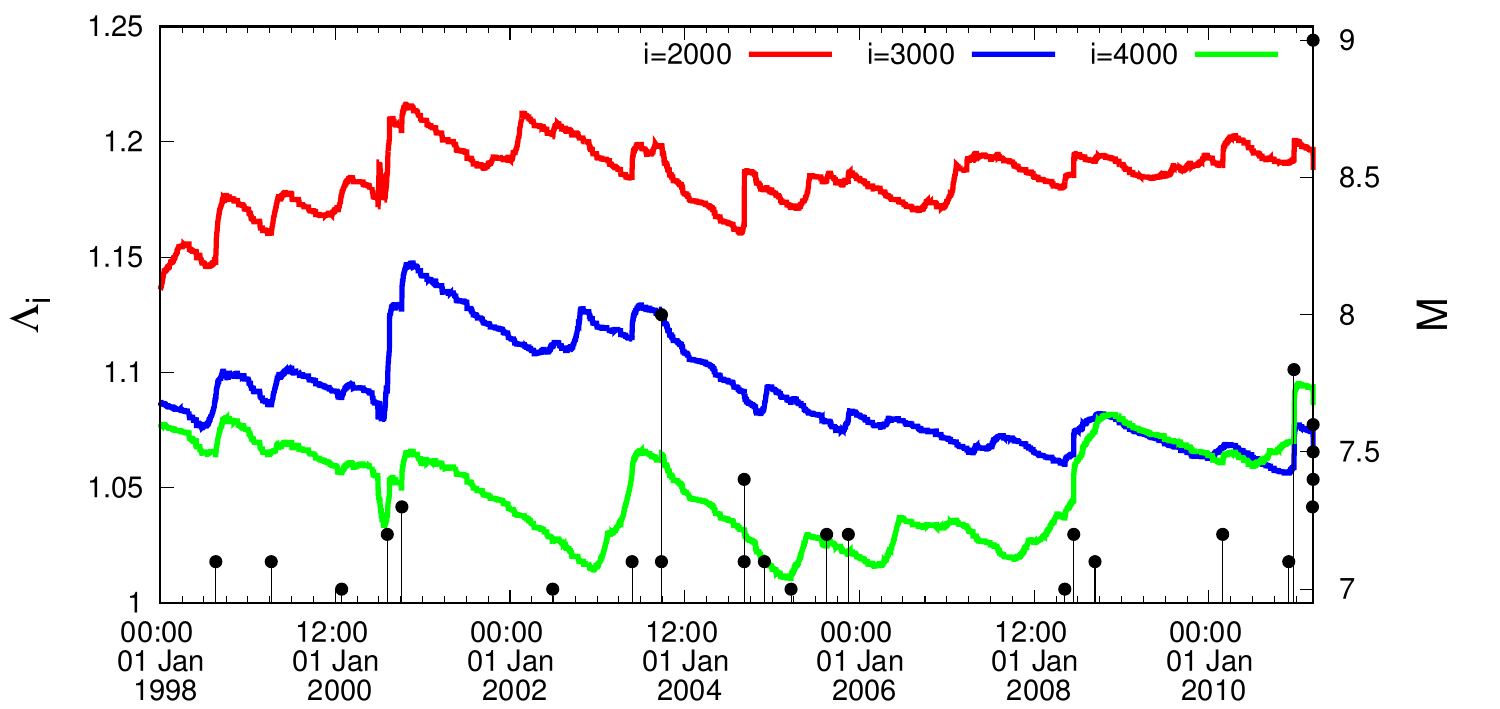}
\caption{The complexity measure $\Lambda_i$ 
for various scales $i=$2000 (red), 3000 (blue), and 4000 (green)
versus the conventional time from 1 January 1998 until the $M$9 Tohoku EQ.}\label{f1}
\end{figure}

 \begin{figure}
\includegraphics[scale=0.35,angle=270]{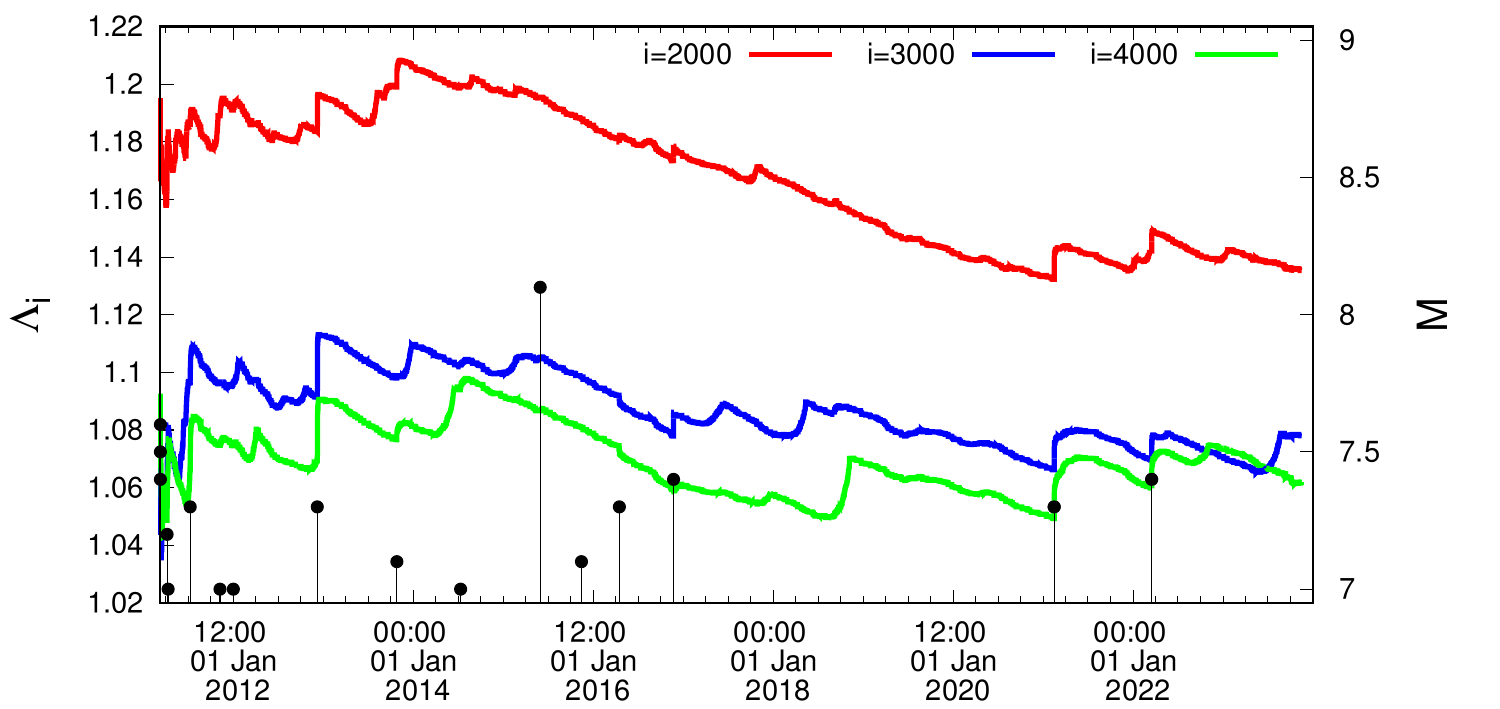}
\caption{The complexity measure $\Lambda_i$ for
 various scales $i=$2000 (red), 3000 (blue), and 4000 (green)
 versus the conventional time from 15:00 LT on 11 March 2011
 until 15 November 2023.
{The strongest EQ during this period 
 is the Ogasawara EQ, see the text.}}\label{f2}
\end{figure}

\begin{figure}
\includegraphics[scale=0.35,angle=270]{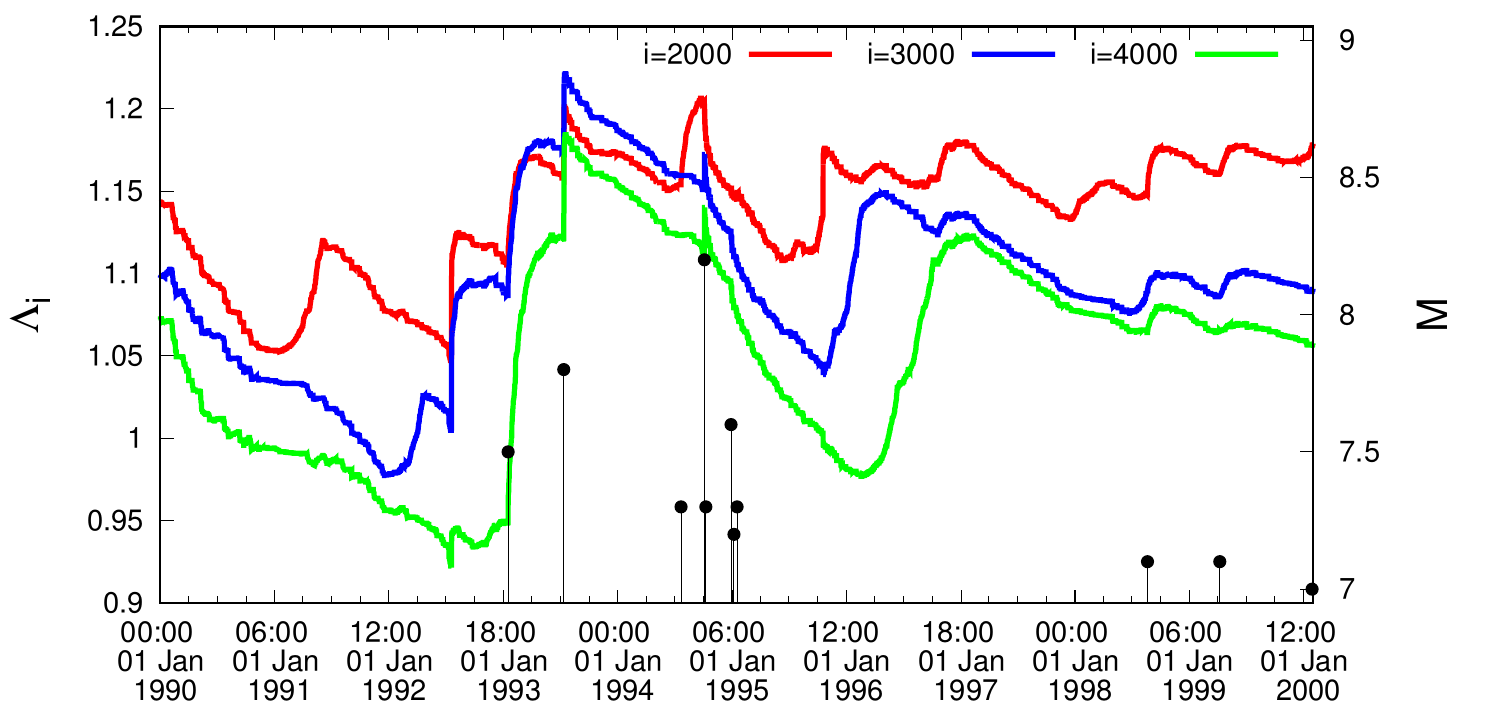}
\caption{The complexity measure $\Lambda_i$ for various scales
 $i=$2000 (red), 3000 (blue), and 4000 (green)
 versus the conventional time from 1 January 1990 until 1 February  2000.}\label{f3}
\end{figure}

The results from Japan concerning the study of $\Lambda_i$ are plotted in Figs.\ref{f1}, \ref{f2}, and \ref{f3}
by starting the computation from 1 January 1984 for the scales $i=$ 2000, 3000, and 4000 events. After a 
careful inspection of these figures the following comments are now in order:

\subsection{Results  from 1 January 1998 until the $M$9 Tohoku EQ occurrence on 11 March 2011} \label{fortoh}

During almost a decade, i.e., during the period from 1 January 1998 until the $M$7.2 EQ on 
14 June 2008, there  exists  no intersection between the 
curves of the three scales $i=2000$, 3000, and 4000 
events since the scale $i=2000$ events lies in 
the highest level, the scale $i=3000$ events in the 
middle level and the scale $i=4000$ events in the lowest level. 
Approximately, from the latter date the curve of the 
scale $i=4000$ events shows a clear increase, 
thus finally almost overlapping the curve of the scale $i=3000$ events 
until almost 5 August 2010. From thereon, however, the curve 
corresponding to $i=4000$ events exceeds the one of 3000 events 
(cf. at this date the two curves intersect) and subsequently it
exhibits an abrupt increase upon the occurrence of the $M$7.8 
EQ on 22 December 2010 in southern Japan at 27.05$^o$N 143.94$^o$E, 
which constitutes an evident intersection. 
Remarkably, on this date (22 December 2010) 
of the abrupt increase of $\Lambda_i$  additional facts 
are observed: The abrupt 
increase conforms to the 
seminal work by Lifshitz and Slyozov \cite{LIF1961} and 
independently by Wagner \cite{WAG1961} (LSW) for phase
 transitions showing that the characteristic size of the 
 minority phase droplets exhibits a scaling behavior 
 in which time ($t$) growth has 
 the form $A (t - t_0)^{1/3}$. 
 It was found that the
 increase $\Delta \Lambda_i $ of $ \Lambda_i $ follows
 the latter form and that the prefactors $A$ are 
 proportional to the scale $i$, while the exponent $(1/3)$ 
is independent of $i$ \cite{ENTROPY18}.   
Furthermore, the Tsallis \cite{TSA88} entropic index $q$ 
exhibits a simultaneous increase 
with the same exponent (1/3) \cite{ENTROPY18}. 
In addition, a minimum $\Delta S_{min}$ of the entropy change $\Delta S$ 
 of seismicity in the entire Japanese region  under time reversal was identified by Sarlis et al. \cite{EPL18}, who also showed 
that the probability to obtain such a minimum by chance is approximately
 3\% thus demonstrating that it is statistically significant. 
 The robustness of the appearance
of $\Delta S_{min}$ on 22 December 2010 upon changing the EQ depth, the EQ magnitude threshold, and the 
size of the area investigated has been documented\citep{EPL18}.
Such a minimum is of precursory nature, 
signaling that a large EQ is impending according to the NTA of the 
Olami-Feder-Christensen (OFC) model for 
earthquakes \cite{OLA92},  which
is probably \cite{RAM06} 
the most studied non-conservative self-organized criticality (SOC) model,
 originated by a simplification 
 of the Burridge and Knopoff spring-block model \cite{BUR67}. 
{ In the OFC model,} NTA showed  that  $\Delta S$ exhibits
 a clear minimum \cite{SPRINGER}  before a large avalanche, which corresponds to a large EQ. 
 Finally, studying the fluctuations $\beta$ of $\kappa_1$ of seismicity in the
entire Japanese region N$_{25}^{46}$E$_{125}^{148}$ 
versus the conventional time
from 1 January 1984 until the Tohoku EQ occurrence on
11 March 2011, we find \cite{EPL19} a large fluctuation of $\beta$ upon the occurrence of the $M$7.8 EQ on 22 December 2010.
 This finding was also
 checked for several scales from i = 150 to 500 events,
which also revealed the following \cite{EPL19}:
upon increasing $i$ it is observed (see Figs. 2b and 4e of Ref. \cite{PNAS13}) that the increase 
$\Delta \beta_i$ of the $\beta_i$ fluctuation on
22 December 2010 becomes distinctly larger – obeying the
interrelation $\Delta \beta_i = 0.5 \ln(i/114.3)$ - 
which does not happen
(see Fig. 4a–d of \cite{PNAS13}) for the increases in the
$\beta$ fluctuations upon the occurrences of all other shallow EQs
in Japan of magnitude 7.6 or larger during the period from
1 January 1984 to the time of the $M$9 Tohoku EQ.
This interrelation
 $\Delta \beta_i = 0.5 \ln(i/114.3)$, see Fig. 2(g)
and (h) of Ref.\cite{EPL19}, has a functional form strikingly reminiscent of
the one discussed by Penrose et al. \cite{Penrose1978} 
in computer simulations of phase separation kinetics using the ideas of
Lifshitz and Slyozov \cite{LIF1961}, see their equation (33) which is
also due to Lifshitz and Slyozov.
 Hence, the $\beta$ fluctuation on 22 December 2010 accompanying the minimum $\Delta S_{min}$ is unique.

\subsection{Results from 15:00 LT on 11 March 2011 until now}\label{now}

During this period, a $M_w$7.9 EQ occurred beneath the 
Ogasawara (Bonin) Islands on 30 May 2015 
 as {depicted} in Fig.\ref{f2}. 
 It occurred at 680 km depth in an area without 
 any known historical seismicity and caused significant
 shaking over a broad
area of Japan at epicentral distances in the
 range 1000--2000 km. 
It
 was the first EQ felt in every Japanese 
 prefecture since intensity observations began in 1884.
This is the deepest
 EQ ever detected 
 (\url{https://www.nationalgeographic.com/science/article/deepest-earthquake-ever-detected-struck-467-miles-beneath-japan}).
and  was also noted \cite{YE16} that globally,
 this is the deepest (680 km centroid depth) event
 with $M_w >$ 7.8 in the seismological records.
 The  Ogasawara  
EQ has not been followed  by an appreciably stronger EQ 
in contrast to the $M$7.8 Chichi-jima  shallow EQ 
which occurred also at Bonin islands at 27.05$^o$N 143.94$^o$E 
on 22 December 2010,  almost three months 
before the $M$9  Tohoku  EQ. 
This could be understood as follows \cite{VAR21}: 
Upon the occurrence of the Chichi-jima  EQ the following
 facts have been observed: First, according
 to Ref. \cite{ENTROPY18} the complexity
 measures $ \Lambda_{2000} $, $ \Lambda_{3000} $ 
 and $ \Lambda_{4000} $, i.e., the $ \Lambda_i $ values 
 at the natural 
 time window lengths (scales) $i$= 2000, 3000 and 4000 events,
 respectively, show 
 a strong abrupt increase $ \Delta \Lambda_i $
in Fig. 7 
 of Ref. \cite{ENTROPY18}  
 on 22 December 2010 and 
 just 
 after the EQ occurrence exhibiting a scaling 
 behavior of the form $\Delta \Lambda_i  = A (t - t_0)^c$ 
 (where the exponent $c$ is 
very close to $1/3$  and $t_0$ is 
approximately 0.2 days after the $M7.8$ 
EQ occurrence), which conforms to LSW.
Second, the order  parameter fluctuations 
showed a unique  change \cite{EPL19}, i.e., 
an increase $ \Delta \beta_i $, which exhibits a functional
 form  consistent 
 with the LSW theory  and the subsequent work 
 of Penrose et al. \cite{Penrose1978} 
 obeying the 
 interrelation $\Delta \beta_i=0.5\ln(i/114.3)$,
 see Fig. 2(g) and (h) of Ref.\cite{EPL19}. 
 Such a behavior has not been observed {along with} the
 occurrence of either the  Ogasawara  EQ 
 or any other shallow EQs in Japan of magnitude 7.6 or larger during 
  the period from 1 January 1984 to the time 
 of the $M$9  Tohoku  EQ \cite{EPL19} (including also the EQ that occurred on 1 January 2024 discussed later).
 
  \begin{figure}
\includegraphics[scale=0.35,angle=270]{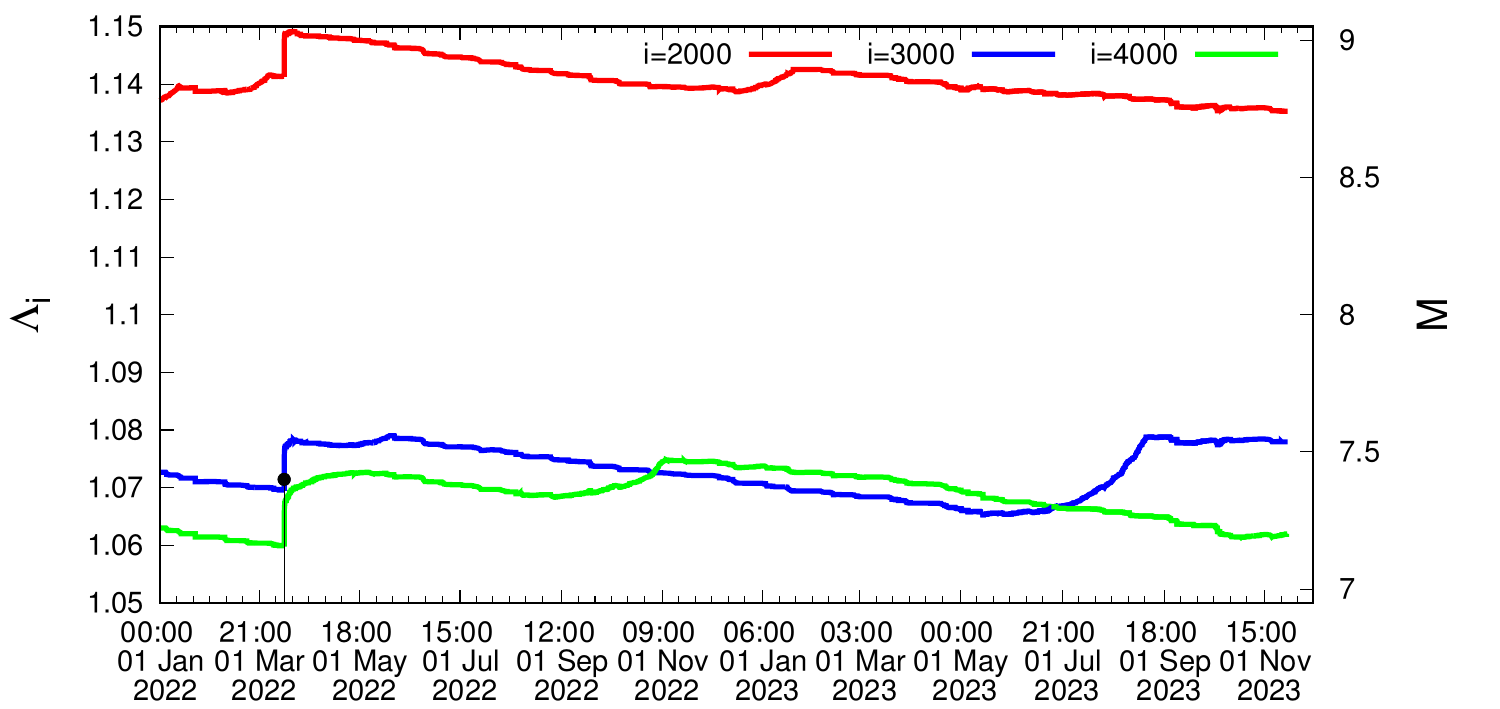}
\caption{The complexity measure $\Lambda_i$ for various
 scales $i=$2000 (red), 3000 (blue), and 4000 (green)
 versus the conventional time from 1 
 January 2022 until 15 November   2023.}\label{f4}
\end{figure} 
 
\begin{figure}
\includegraphics[scale=0.52,angle=0]{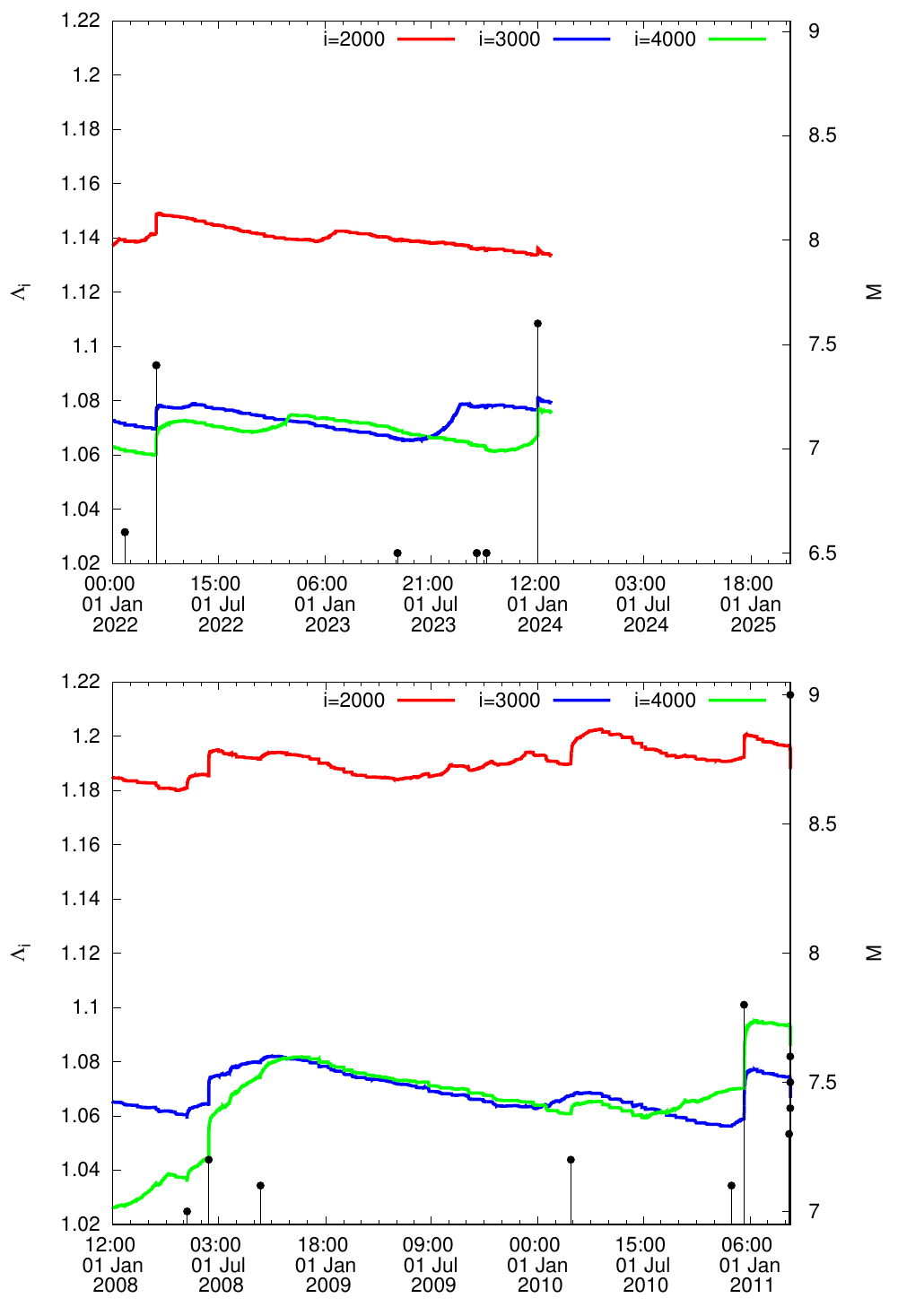}
{\caption{The complexity measure $\Lambda_i$ 
for various scales $i=2000$ (red), 3000 (blue),
and 4000 (green) before the $M$7.6 EQ on the 
west coast of Japan on 1 January 2024 (upper panel) and the $M$9 Tohoku EQ on 11 March 2011 (lower panel).}\label{f5}}
\end{figure}

 An additional important fact is the following: 
 On 27 October  2022, the curve corresponding 
 to the scale $i=4000$ events (green) intersects the 
 one for the scale $i=3000$ (blue), 
but the latter on 27 June 2023  recovers 
(see Fig.\ref{f4}).
 This phenomenon has been followed very carefully
-since it started as described in Ref. \cite{VAR23J}- compared to the one 
that preceded the $M$9 Tohoku EQ (Fig. \ref{f5}).  

\subsection{Results from 1 January 1990 until 1 February 2000}
In the relevant plot (Fig.\ref{f3}), we observe 
that mostly the curve corresponding to 
the scale $i=2000$ events lies in the 
highest level, the curve $i=3000$ events 
in the middle and the curve $i=4000$ events 
in the lowest level. There exists, however, the following interesting 
intersection: Around 8 March 1993 the
 curve $i=3000$ events jumps to the highest 
 level and remains so until 24 July 1994; 
 subsequently the curve $i=2000$ events 
 returns to the highest level and 
 after that a $M$8.2 EQ occurs on 4 October 1994.
 This is an additional case where a major 
 EQ happens after the detection of an intersection of  $\Lambda_i$ curves.
 
 \section{Discussion}
 
 An EQ of  JMA magnitude $M$=7.6  (USGS reported $M_w=7.5$,
  see, e.g., \url{https://earthquake.usgs.gov/earthquakes/eventpage/us6000m0xl})
   with epicenter at 37.50$^o$N 137.27$^o$E occurred on the west coast of Japan on 1
  January 2024, i.e., almost 
3$\frac{1}{2}$  weeks after drawing attention in Ref.\cite{VAR23J} to the important 
  fact focused on the phenomenon described in the last 7 lines 
  of Section III.B along with Fig.\ref{f4}. Referring to the intersection 
mentioned there, i.e., the curve corresponding to the scale $i$=4000 events 
(green) intersects the one for the scale $i$=3000 (blue), the following comments 
are now in order: First, the two EQs of magnitude close to $M$8, i.e., 
the 2003 Tokachi EQ (see Fig.\ref{f1}) and the 2015 
Ogasawara EQ (see Fig.\ref{f2}), have not been preceded by an intersection 
(see also Fig.\ref{f2} where the green curve approaches -but not 
intersects- the blue curve). Second, concerning the $M$8.2 EQ in 1994 -exceeding 
the aforementioned two EQs of magnitude close to $M$8- there exists
an intersection, however, since the curve $i=3000$ events in Fig.\ref{f3}
jumps to the highest level and an intersection occurs with the curve 
$i=2000$ events (red) around 8 March 1993. 
In other words, before 27 October 2022 the only intersection 
 between the curves corresponding to the  scales 
$i=4000$ and $i=3000$ events was observed
before the Tohoku $M$9 EQ, see Section III.A. Thus
 the phenomenon emerged in Fig.\ref{f4} and mentioned in the last lines of 
Section III.B has only appeared before the Tohoku $M$9 EQ 
as can be visualized 
in the lower panel of Fig.\ref{f5} -which is just an excerpt of Fig.\ref{f1}- showing
 the following sequence before the Tohoku mainshock:
(a)for several months (i.e., approximately 14.5 months
 from 25 October 2008 to 10 January 2010) the 
$i=4000$ events curve slightly exceeded the $i=3000$ curve 
(which actually occurred in the aforementioned 2023 case)
and  
(b)subsequently the $i=3000$ curve recovered for approximately 7 months
(from 10 January 2010 to 5 August 2010). Then, a clear intersection
occurs on around 5 August 2010 and the $i=4000$ events curve starts 
to increase more rapidly until 22 December 2010 when a $M$7.8 EQ 
occurred. Almost two weeks later an SES activity started (with 
a duration of around 10 days) and almost two months later the $M$9 Tohoku 
mainshock occurred. In short, the {aforementioned} comments shed more light 
on why the   phenomenon in 2023
-depicted in Fig.\ref{f4} and Fig.\ref{f5} (upper panel)- has been,   followed very carefully 
as mentioned in Ref.\cite{VAR23J}
by comparing to the one that preceded the $M$9 Tohoku EQ.

{We now proceed to the estimation of the statistical significance
 of the observed phenomenon. As mentioned above on 8 March 1993, i.e., 
 19 months before 
the East-Off Hokaido $M$8.2 EQ on 4 October 1994, $\Lambda_{3000}$ exceeded
 $\Lambda_{2000}$ for the first time. A similar phenomenon concerning 
$\Lambda_{4000}$ exceeding $\Lambda_{3000}$ occurred on 14 June 2008, 
i.e., 32 months before the $M$9 Tohoku EQ on 11 March 2011, see Section \ref{fortoh}.
Thus, assuming that an alarm is set ON when such intersections occur,
 we find that for the time period from 1 January 1990 to 1 January 2022 
 consisting of 384 months 
the probability to have the alarm ON is $p_{ON}=(19+32)/384=13.28$\%.
 Obviously, the $p$-value to hit by chance both EQs of magnitude
 $M$8.2 or larger  is  $p=p_{ON}^2=1.76$\%, which points
 to statistical significance of the phenomenon observed.
 We clarify that the present calculation of the statistical significance does 
 not include the period depicted in Fig.\ref{f4} because  
 the intersection displayed after 1 January 2022 is still under investigation,
 as already mentioned in Section \ref{now}.}

\section{Summary and Conclusions}
Let us summarize: $\Lambda_i$ is a complexity measure in NTA quantifying 
the fluctuations of the entropy change $\Delta S_i$ under time-reversal.
Studying the evolution of  $\Lambda_i$ curves for the 
seismicity of Japan during the 
last 39 years for various scales $i(=$2000 to 4000 events), we find
that intersections of these curves occurred before the two strongest EQs 
(exceeding $M$8), i.e., the $M$9 Tohoku EQ on 11 March 2011 and the 
East-Off Hokaido $M$8.2 EQ on 4 October 1994. The same phenomenon 
is ascertained  
(by an inspection of Fig. 8.17(a) of Ref.\cite{SPRINGER23})
before the deadly Chiapas $M$8.2 EQ,
 which is Mexico's largest EQ in more 
than a century.

\providecommand{\noopsort}[1]{}\providecommand{\singleletter}[1]{#1}%

\end{document}